\documentclass[twocolumn,showpacs,preprintnumbers,amsmath,amssymb]{revtex4}

\usepackage{graphicx}
\usepackage{bm}% bold math

\begin{document}

\preprint{APS/}

\title{Magnetization reversal by injection and transfer of spin: experiments and theory}

\author{A. Fert, V. Cros, J.-M.
George, J. Grollier, H. Jaffr\`es} \affiliation{Unit\'e Mixte de
Physique CNRS/THALES, Domaine de Corbeville, 91404 Orsay, and
Universit\'e Paris-Sud, 91405 Orsay, France}
\author{A. Hamzi\'{c}}
 \altaffiliation[~on leave from ] {the Department of Physics,
Faculty of Science, HR-10002 Zagreb, Croatia.}
 \affiliation{Unit\'e Mixte de Physique CNRS/THALES, Domaine de
Corbeville, 91404 Orsay, and Universit\'e Paris-Sud, 91405 Orsay,
France}
\author{A. Vaur\`es}
\affiliation{Unit\'e Mixte de Physique CNRS/THALES, Domaine de
Corbeville, 91404 Orsay, and Universit\'e Paris-Sud, 91405 Orsay,
France}
\author{G. Faini}
\affiliation{Laboratoire de Photonique et de Nanostructures,
LPN-CNRS, Route de Nozay, 91460 Marcoussis, France}
\author{J. Ben Youssef, H. Le Gall}
\affiliation{Laboratoire de Magn\'etisme de Bretagne, 29285 Brest,
France}

\date{\today}

\begin{abstract}
Reversing the magnetization of a ferromagnet by spin transfer from
a current, rather than by applying a magnetic field, is the
central idea of an extensive current research. After a review of
our experiments of current-induced magnetization reversal in
Co/Cu/Co trilayered pillars, we present the model we have worked
out for the calculation of the current-induced torque and the
interpretation of the experiments.
\end{abstract}

\maketitle

The concept of magnetization reversal by spin transfer from a
spin-polarized current was introduced in 1996 by Slonczewski~[1].
Similar ideas of spin transfer had also appeared in the earlier
work of Berger~[2] on current-induced domain wall motion.
Convincing experiments of magnetization reversal by spin transfer
on pillar-shaped multilayers~[3-6], nanowires~[7] or
nanocontacts~[8] have been recently performed and several
theoretical approaches, extending the initial theory, have  also
been developed~[9-19]. From the application point of view,
magnetization reversal by spin transfer can be of great interest
to switch spintronic devices (MRAM for example), especially if the
required current density - presently around $10^{7}$ A/cm$^{2}$ -
can be reduced by approximately  an order of magnitude.

We present a summary of our experiments on Co/Cu/Co pillars,
describe a calculation model for the critical currents as a
function of - mainly - CPP-GMR data and we discuss its application
to experiments.

\vspace{0.1in} \textbf{I. Experiments} \vspace{0.1in}

We present experiments on pillar-shaped Co1(2.5 nm)/Cu(10
nm)/Co2(15 nm) trilayers.~The submicronic (200 $\times$ 600
nm$^{2}$) pillars are fabricated by e-beam lithography~[5]. The
CCP-GMR of the trilayer is used to detect the changes of the
magnetic configuration (the difference between the resistances of
the P and AP configurations is about 1~m$\Omega$). For all the
experiments we describe, the initial magnetic configuration is a
parallel (P) one, with the magnetic moments of the Co layers along
the positive direction of an axis parallel to the long side of the
rectangular pillar. A field $H_{\mathrm{appl}}$ is applied along
the positive direction of this axis (thus stabilizing this initial
P magnetic configuration). We record the variation of the
resistance ($R$) as the current $I$ is increased or decreased
(positive $I$ means electrons going from the thick Co layer to the
thin one). The results we report here are obtained at 30 K (the
critical currents are smaller at room temperature).

\begin{figure}
\begin{center}
\includegraphics*[width=7.0cm]{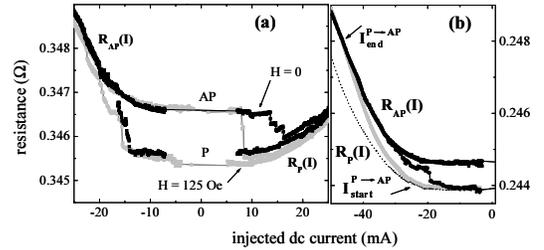}
\end{center}
\caption{Resistance \textit{vs.} dc current: (a) sample 1 for
$H_{\mathrm{appl}}$ = 0 (black) and $H_{\mathrm{appl}}$ = 125~Oe
(grey); (b) sample 2 for $H_{\mathrm{appl}}$ = 0 (black),
$H_{\mathrm{appl}}$ = + 500~Oe (grey) and $H_{\mathrm{appl}}$ = +
5000~Oe (dotted line).}\label{fig1}
\end{figure}

In Fig. 1(a), we present a typical variation of the resistance $R$
as a function of the current, for $H_{\mathrm{appl}}$ = 0 and
+~125~Oe. Starting from a P configuration at $I=0$ and increasing
the current to positive values, we observe only a small
progressive and reversible increase of the resistance, which can
be ascribed to Joule heating (this has also been  seen in all
other experiments on pillars~[3-6] when the current density
reaches the range of $10^{7}$ A/cm$^{2}$). In contrast, when the
current is negative and at a critical value $I_{C}^{P\rightarrow
AP}$, there is an irreversible jump of the resistance ($\Delta R
\approx$ 1 m$\Omega$), which corresponds to a transition from the
P to the AP configuration (reversal of the magnetic moment of the
thin Co layer). The trilayer then remains in this high resistance
state (the $R_{AP}(I)$ curve) until the current is reversed and
increased to the critical value $I_{C}^{AP\rightarrow P}$, where
the resistance drops back to the $R_{P}(I)$ curve. This type of
hysteretic $R(I)$ cycle is characteristic of the magnetization
reversal by spin injection in \textit{regime A}.

For $H_{\mathrm{appl}}$ = 0, $I_{C}^{P\rightarrow AP}\cong -$
15~mA (current density $j_{C}^{P\rightarrow AP}\cong - ~1.25\times
10^{7}$ A/cm$^{2}$) and $I_{C}^{AP\rightarrow P}\cong$ +~14~mA
($j_{C}^{AP\rightarrow P}\cong + ~1.17\times 10^{7}$ A/cm$^{2}$).
A positive field, which stabilizes the P configuration, shifts
slightly the critical currents;  $|I_{C}^{P\rightarrow AP}|$
increases and  $I_{C}^{AP\rightarrow P}$  decreases (note that the
relatively larger shift of $I_{C}^{AP\rightarrow P}$ at 125~Oe in
Fig.~1(a) is specific to the approach to the crossover to
\textit{regime B} at about 150~Oe).

The $R(I)$ curve for $H_{\mathrm{appl}}$ = + 500~Oe, shown in Fig.
1(b), illustrates the different behavior when the applied field is
higher (\textit{regime B}). Starting from $I=0$ in a P
configuration (on the $R_{P}(I)$ curve), a large enough negative
current still induces a transition from P to AP, but now this
transition is \textit{progressive} and \textit{reversible}. The
$R(I)$ curve departs from the $R_{P}(I)$ curve at
$I_{start}^{P\rightarrow AP}\cong -~25$~mA
($j_{start}^{P\rightarrow AP}\cong -~2.08\times 10^{7}$
A/cm$^{2}$) and catches up the $R_{AP}(I)$ curve only at
$I_{end}^{P\rightarrow AP}\cong - $~45~mA ($j_{end}^{P\rightarrow
AP}\cong - ~3.75\times 10^{7}$ A/cm$^{2}$). On the way back,
reversibly, $R(I)$ departs from $R_{AP}(I)$ at
$I_{start}^{AP\rightarrow P}=I_{end}^{P\rightarrow AP}\cong -~
$45~mA and reaches finally $R_{P}(I)$ at $I_{end}^{AP\rightarrow
P}=I_{start}^{P\rightarrow AP}\cong - $~25~mA. At higher field,
the transition is similarly progressive and reversible, but occurs
in a higher negative current range. Finally, for very large
applied field ($H_{\mathrm{appl}}$ = 5000~Oe), the transition is
out of our experimental current range, and the recorded curve is
simply $R_{P}(I)$.

\begin{figure}
\centering
\includegraphics[keepaspectratio=1,width=7.0cm]{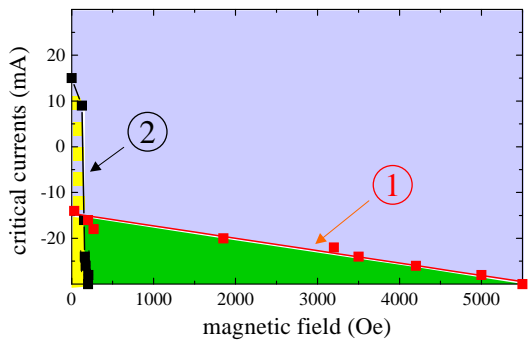}
\caption{Instability lines of the P and AP configurations (sample
1). The P configuration is stable above \textit{line 1} and
unstable below. The AP one is stable below \textit{line 2} and
unstable above. At low field (\textit{regime A}), the stability
zones of P (blue) and AP (yellow) overlap between \textit{lines 1}
and \textit{2} (stripes). At high field (\textit{regime B}), there
is a zone (green) between \textit{lines 1} and \textit{2} where
none of the P and AP configurations is stable. Equations of
\textit{lines 1} and \textit{2} are derived from a LLG equation
for uniaxial anisotropy $H_{\mathrm{an}}$~[18]. The magnetic field
includes $H_{\mathrm{appl.}}$ and, possibly, interlayer coupling
fields. \textit{lines 1} and \textit{2} cross at about
$H_{\mathrm{an}}$.}\label{fig2}
\end{figure}

The experimental results presented above can be summarized by the
diagram of Fig.2. This type of diagram is obtained~[18] by
introducing the current-induced torque into a
Landau-Lifshitz-Gilbert (LLG) motion equation to study the
stability/instability of the moment of the magnetic thin layer
(the moment of the thick layer supposed being pinned). The P
configuration is stable above \textit{line 1} and unstable below.
The AP configuration is stable below \textit{line 2} and unstable
above.

\textit{Regime A} corresponds to $H_{\mathrm{appl}}$ smaller than
the field at which \textit{line 2} crosses \textit{line 1}. In
this regime, there is an overlap between the stability regions of
P and AP. Starting from a P configuration at zero current and
moving downward on a vertical line, the P configuration becomes
unstable at the negative current $I_{C}^{P\rightarrow AP}$
corresponding to the crossing point with \textit{line 1}. As this
point in the stability region of the AP configuration, the
unstable P configuration can switch directly to the stable AP
configuration. On the way back, the AP configuration remains
stable until the crossing point with \textit{line 2} at
$I_{C}^{AP\rightarrow P}$ (positive), where it can switch directly
to a stable P configuration. This accounts for the direct
transitions and hysteretic behavior of \textit{regime A} in
Fig.1(a).

In \textit{regime B}, for $H_{\mathrm{appl}}$ above the crossing
point of \textit{lines 1} and \textit{2}, none of the P and AP
configurations is stable in the region between \textit{lines 1}
and \textit{2}. Going down along a vertical line, the P
configuration becomes unstable at the crossing point with
\textit{line 1} ($I_{start}^{P\rightarrow AP}$) and the system
departs from this configuration. But the AP configuration is still
unstable at this current and can be reached only at the crossing
point with \textit{line 2} ($I_{end}^{P\rightarrow AP}$). On the
way back, reversibly, the AP configuration becomes unstable at the
crossing point with \textit{line 2} ($I_{start}^{AP\rightarrow
P}=I_{end}^{P\rightarrow AP}$), but a stable P configuration is
reached only at the crossing point with \textit{line 1}
($I_{end}^{AP\rightarrow P}=I_{start}^{P\rightarrow AP}$). This
accounts for the behavior of Fig.~1(b). The state of the system
during the progressive transition between P and AP can be
described as a state of maintained precession.

The critical lines of the diagram of Fig.2 can also be derived
from the variation of $R$ along a horizontal line, for example
from the $R(H_{\mathrm{appl}})$ curves of Fig.3 for sample 2. The
$R(H_{\mathrm{appl}})$ curve for $I = + 50$~mA is flat, i.e. there
is no GMR. This is because, along an horizontal line in the upper
part of the diagram of Fig.2, the P configuration is always
stable. For negative current, on the other hand, the
$R(H_{\mathrm{appl}})$ curves mimic the GMR curves of an
antiferromagnetically coupled trilayer, in which the
antiferromagnetic coupling would increase when the current becomes
more negative.
\begin{figure}
\centering
\includegraphics[keepaspectratio=1,width=7.0cm]{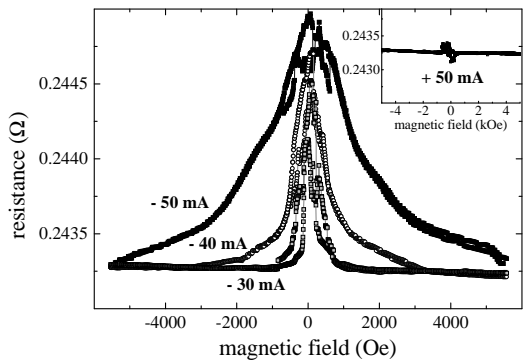}
\caption{Resistance \textit{vs.} applied magnetic field in sample
2 for $I= -$~50~mA, $-~40$ mA, and $-$~30 mA. For clarity, the
curves have been shifted vertically to have the same high field
baseline. inset :$R(H)$ for $I = + ~50$~mA.}\label{fig3}
\end{figure}
This can be expected from the diagram of Fig.2. For example,
starting from high field at $I = -$~50~mA, the upturn from the
baseline at about $H_{\mathrm{appl}}$ = +~5600~Oe indicates the
beginning of the progressive transition from P to AP at the
crossing point with \textit{line 1}. As $H_{\mathrm{appl}}$ is
decreased further, the progressive (and reversible) increase of
$R$ reflects the progressive crossover from P to AP on a
horizontal line between \textit{line 1} at 5600~Oe and
\textit{line 2} at a field in the range 100$-$200~Oe. When the
moment of the thick Co layer is reversed in a small negative
field, the P configuration being unstable and the AP stable in
this region of the diagram, the moment of the thin layer is also
reversed to restore the AP configuration, so that $R$ is
practically not affected by the coupled reversal of both layers.

We conclude that the main features of the experimental results fit
into the frame of the diagram of Fig.~2. In Section IV, we discus
more quantitatively the influence of parameters such as layer
thicknesses, spin diffusion length, etc. The final remark of this
Section is that the phase diagram of Fig.~2 comes from an
oversimplified model assuming that the only current-induced
excitations are precessions of a global magnetization vector due
to transverse spin transfer. Several types of additional effects
can be expected from non-uniform precessions, or, more generally,
from other modes of current-induced excitations. For example,
excitation of magnons is probably a significant dissipation
mechanism in the stage of maintained precession and also a
dissipation channel of the longitudinal spin accumulation at high
current density. These additional excitation modes should also be
reflected in the resistivity and probably account for some
features of the experimental results~[3-8] which are not described
by the scheme of Fig.~2 for pure rotations. Others effects~[6] are
also expected from exchange or dipolar interlayer couplings which
can play the same role as the applied field in Fig.2.

\vspace{0.1in} \textbf{II. Theoretical Model} \vspace{0.1in}

The magnetization of a magnetic layer can be reversed by spin
transfer if the spin polarization of the injected current and the
magnetization of the layer are non-colinear. In a multilayered
structure this requires a non-colinear configuration of the
magnetizations of the different layers. The  transfer from an
obliquely polarized spin current running into a magnetic layer is
associated with the alignment of the polarization of the current
inside the layer along the magnetization axis. If the
current-layer interaction is spin conserving (exchange-like), this
implies that the transverse component of the spin current is
absorbed and transferred to the layer. This is the spin transfer
concept introduced by Slonczewski~[1]. The contribution of this
transfer to the motion equation of the total spin \textbf{S} of
the layer is written as:
\begin{equation}
(d\textbf{S}/dt)_{j}=absorbed~transv.~spin~current \label{torque1}
\end{equation}
or, in other words, a torque equal to the absorbed spin current
multiplied by $\hbar$ is acting on the magnetic moment of the
layer.

Several mechanisms contribute to the transfer of the transverse
component of a spin current running into a magnetic layer~[12].
First, due to the spin dependence of the reflection/transmission
process at the interface with a ferromagnet, the transverse
component is reduced and rotated in the transmitted spin current.
What  remains of transverse component then disappears (is
transferred) by incoherent precession of the electron spins in the
exchange field of the ferromagnet. Ab-initio calculations ~[12]
show that, for a metal like Co, the transverse spin current is
almost completely absorbed at a distance of the order of 1 nm from
the interface. In these conditions, the spin transfer is a
quasi-interfacial effect and, in our calculation, is expressed by
interface boundary conditions (in the same way as interface
resistances are introduced in boundary conditions for the theory
of CPP-GMR~[20]). On the other hand, the longitudinal component of
the spin current in the magnetic layers and all its components in
the nonmagnetic layers vary at the much longer scale of the spin
diffusion length $l_{\mathrm{sf}}$ (60 nm in Co, about 1 $\mu$m in
Cu). They can be calculated by solving diffusive transport
equations for the entire structure, as in the theory of the
CPP-GMR. An essential point is that, for a non-colinear
configuration with different orientations of the longitudinal axes
in different layers, the longitudinal and transverse components of
the spin current are inter-twined from one layer to the next one,
so that a global solution for both the longitudinal and transverse
component and for the entire structure is required.

The calculation of our model can be summarized as follows. We
consider a N$_{L}$/F$_{1}$/N/F$_{2}$/N$_{R}$ structure, where
F$_{1}$ (thin) and F$_{2}$ (thick) are ferromagnetic layers
separated by a $t_{N}$ thick nonmagnetic layer N. N$_{L}$ and
N$_{R}$ are two semi-infinite nonmagnetic layers (leads). For
simplicity we assume that F$_{1}$ and F$_{2}$ (N, N$_{L}$ and
N$_{R}$) are made of the same ferromagnetic (nonmagnetic)
material. The current is along the $x$ axis  perpendicular to the
layers. $\widehat{m}$(x) and $\widehat{j}$(x) are the $2\times 2$
matrices representing respectively  the spin accumulation and the
current density:

\begin{eqnarray}
\widehat{j}(x) & = & \frac{j_{e}}{e}\widehat{I} +
j_{m,x}(x)\widehat{\sigma}_{x}+ j_{m,y}(x)\widehat{\sigma}_{y} +
j_{m,z}(x)\widehat{\sigma}_{z} \nonumber\\
\widehat{m}(x) & = & m_{x}(x)\widehat{\sigma}_{x} +
m_{y}(x)\widehat{\sigma}_{y} + m_{z}(x)\widehat{\sigma}_{z}
 \label{accu-curr}
\end{eqnarray}

where $\widehat{\sigma}_{x}$, $\widehat{\sigma}_{y}$ and
$\widehat{\sigma}_{z}$ are the three Pauli matrices and
$\widehat{I}$ is the unitary matrix. Spin accumulation and current
are defined as in Ref [13]. If we call $z_{i}$ the local spin
polarization axis ($z_{i} = z_{1}$ in F$_{1}$, $z_{i} = z_{2}$ in
F$_{2}$), $m_{z_{i}}$ ($j_{m,z_{i}}$) is the longitudinal
component of the spin accumulation vector \textbf{m} (spin current
vector \textbf{j$_{m}$}), $m_{x_{i}}$ and $m_{y_{i}}$
($j_{m,x_{i}}$ and $j_{m,y_{i}}$) are the transverse components of
\textbf{m} (\textbf{j$_{m}$}).

To derive the critical currents for the instability of the P and
AP configurations, we need only to calculate the current-induced
torque in the simple limit where the angle between the
magnetizations of the magnetic layers is small or close to $\pi$
($\theta$ or $\pi-\theta$, with $\theta$ small). The first step,
before introducing the small angle $\theta$, is the calculation of
the longitudinal spin current $j_{mz}$ and spin accumulation
$m_{z}$ in a colinear configuration ($\theta=0$). This is done by
using the standard diffusive transport equations of the theory of
the CPP-GMR with parameters (spin dependent interface resistances,
interface spin memory loss coefficient, spin diffusion lengths,
etc) derived from CPP-GMR experiments~[21,22]. An example of the
result for the P configuration of a Co/Cu/Co trilayer is shown at
the top left of Fig.~4. In the bottom part of Fig.~4, we represent
the situation when a small deviation $\theta$ from the parallel
colinear configuration above is introduced. The spin accumulation
in the Cu spacer layer is a constant vector \textbf{m} (as,
generally,
$t_{\mathrm{Cu}}$~$\ll$~$l_{\mathrm{sf}}^{\mathrm{Cu}}$). With
respect to the colinear configuration, the amplitude of \textbf{m}
has changed by a quantity of the first order in $\theta$ (we omit
this part of the calculation). However, to calculate the torque at
first order in $\theta$, we can neglect this change and assume
$|\mathbf{m}|$ = $m^{P}_{Cu}$, where $m^{P}_{Cu}$ is the spin
accumulation $m_{z}$ in Cu for the P colinear configuration. On
the other hand, \textbf{m} cannot be parallel to both $z_{1}$ and
$z_{2}$, and its orientation in the frame of the thin layer is
characterized by the unknown angles $\theta_{m}$ (of the order of
$\theta$) and $\chi$. These angles will be determined later by
self-consistency conditions for the whole structure. The key
point, explaining the injection of a large transverse spin current
into the thin magnetic layer, is the discontinuity of transverse
spin accumulation between the two sides of the interface between
Cu and Co1, $|\mathbf{m}|$ = $\theta_{m}m^{P}_{Cu}$ in Cu and
$|\mathbf{m}|$ = 0 in Co1. This is equivalent to a huge gradient
of spin accumulation and generates a large transverse spin
diffusion current running into the interface where it is absorbed
or reflected. A straightforward angular integration, illustrated
at the top right of Fig.~4, gives for the incoming transverse spin
current:
\begin{equation}
j_{m,\perp}^{inc.}=\frac{1}{4}\theta_{m}e^{i\chi}m^{P}_{Cu}v_{F}\label{currperp}
\end{equation}
where $j_{m,\perp}^{inc.}=j_{m,x}^{inc.}$ + i$j_{m,y}^{inc.}$ and
$v_{F}$ is the Fermi velocity. Eq.(\ref{currperp}) holds for a
spacer thickness of the order of the mean free path or larger. A
part of this incoming transverse spin current is reflected into Cu
at the Cu/Co1 interface. The remaining part absorbed in the
interfacial precession zone can be written as
$j_{m,\perp}^{abs.}=te^{i\epsilon}j_{m,\perp}^{inc.}$, where the
coefficient $t$ and the rotation angle $\epsilon$ have been
calculated [12] for a large number of interfaces. This leads to:
\begin{equation}
j_{m,\perp}^{abs.}=\frac{1}{4}\theta_{m}te^{i(\chi+\epsilon)}m^{P}_{Cu}v_{F}\label{currabs}
\end{equation}
For thinner spacer layers, a contribution to the diffusion current
comes also from the thick magnetic layer and $j_{m,\perp}^{abs.}$
includes an additional term in $m^{P}_{Co}$~[16]. The scale of the
transverse spin current of Eq.(\ref{currabs}) is the product
$m^{P}_{Cu}v_{F}$ (or $m^{AP}_{Cu}v_{F}$ around the AP state),
where $m^{P}_{Cu}$ is controlled by the spin relaxation in the
system. $m^{P}_{Cu}v_{F}$ is of the order of $(j_{e}/e) \langle
l_{\mathrm{sf}}/\lambda \rangle$, where $ \langle
l_{\mathrm{sf}}/\lambda \rangle$ is a mean value of the ratio of
the spin diffusion length (SDL) to the mean free path (MFP) in the
structure (including the leads), and can be definitely larger than
the charge current $j_{e}/e$. In most cases, the transverse spin
current of Eq.(\ref{currabs}) will be larger than the current
$\theta_{m}j^{P(AP)}_{m,Cu}$ corresponding to the projection of
the longitudinal spin current in the colinear configuration (the
diffusion spin current coming from the gradient of spin
accumulation).

The unknown angles $\theta_{m}$ and $\chi$ are calculated~[16] by
imposing a global cancellation of the transverse spin currents
outgoing from or reflected into the spacer layer. In the case of a
small deviation $\theta$ from the P configuration, for example,
this leads to $\theta_{m}=\theta/2$ and $\chi=\pi/2$, and finally,
from Eq.(\ref{torque1}), to the following general expression of
the torque $\mathbf{\Gamma}^{P}$:
\begin{eqnarray}
\frac{\mathbf{\Gamma^{P}}}{\hbar}=[(\frac{v_{F}m^{P}_{Cu}}{8}+\frac{j^{P}_{m,Cu}}{2})(1-e^{-t_{Cu}/\lambda_{Cu}}) \nonumber\\
+(\frac{v_{F}m^{P}_{Co}}{4}+j^{P}_{m,Co})e^{-t_{Cu}/\lambda_{Cu}}]
\nonumber \\\times \mathbf{M_{1}\wedge (M_{1}\wedge
M_{2})}\label{torque2}
\end{eqnarray}
with a similar expression for $\mathbf{\Gamma}^{AP}$
($\mathbf{M_{1}}$ and $\mathbf{M_{2}}$ are unit vectors along the
magnetizations, $m_{Co}^{P}$ and $j_{Co}^{P}$ are the spin
accumulation and current at the Cu/Co2 interface in the colinear
configuration). As ab-initio calculations  have shown that, for
most interfaces between classical magnetic and nonmagnetic metals
[12], $t$ is always close to 1 and $\epsilon$ very small ($t\cong
0.92$ and $\epsilon$ smaller than $3\times 10^{-2}$ for
Cu(111)/Co, for example), we have supposed $t=1$, $\epsilon=0$ and
kept only the term $\mathbf{M_{1}\wedge (M_{1}\wedge M_{2})}$ in
an expression of the form $\mathbf{[\cos(\epsilon)M_{1}\wedge
(M_{1}\wedge M_{2}) + \sin(\epsilon)M_{1}\wedge M_{2}]}$ (assuming
$\epsilon=0$ is equivalent to neglecting the small imaginary parts
of the mixing conductances in circuit theory~[15]). In.
Eq.(\ref{torque2}) we have also neglected the interfacial memory
loss of transverse spin by spin-orbit effects (the longitudinal
spin memory loss at the interfaces~[21] is already taken into
account in the calculation of \textbf{m} and \textbf{j}$_{m}$ in
the colinear configuration).

\begin{figure} \centering
\includegraphics[keepaspectratio=1,width=7.0cm]{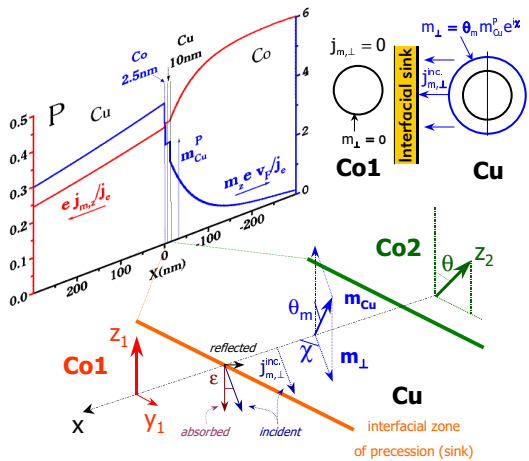}
\caption{Top left: Profile of the spin current \textbf{j}$_{m,z}$
and spin accumulation \textbf{m}$_{z}$ calculated from diffusive
CPP-transport equations and CPP-GMR data for a (Cu/Co1~2.5 nm/Cu~
10~nm/Co2~$\infty$) structure in a parallel colinear configuration
with an electron current ($j_{e}$) going to the left. Bottom: For
a small angle $\theta$ between the polarization axes $z_{1}$
(vertical) and $z_{2}$ of the same structure, 3D sketch
representing the spin accumulation \textbf{m} in the Cu layer
($\mathbf{|m|}=m^{P}_{Cu}$ of the colinear configuration), its
transverse component $m_{\perp}$ and the transverse component of
the induced spin currents diffusing to, reflected from and
absorbed by the Co1 layer. The angles $\theta_{m}$ and $\chi_{m}$
characterize the orientation of the vector \textbf{m} in the frame
of Co1. Top right: Schematic illustrating the calculation of the
transverse spin diffusion current generated by the transverse spin
accumulation on the Cu side of the Co1/Cu interface.}\label{fig4}
\end{figure}

The important feature in Eq.(\ref{torque2}) is the relation of the
torque at small angle to the spin accumulation \textbf{m} and spin
current \textbf{j$_{m}$} calculated for the P and AP colinear
configurations. We emphasize that, due to the relevant length
scale of this calculation, the result for $\mathbf{\Gamma}$
involves the entire structure (including a length of the order of
the SDL in the leads). The spin currents $j^{P(AP)}_{m,Cu}$ are
only a fraction of the charge current $j_{e}/e$. In contrast the
terms $v_{F}m^{P(AP)}$, reflecting the diffusion currents
generated by the transverse spin accumulation discontinuities in a
non-colinear system, are of the order of $(j_{e}/e) \langle
l_{\mathrm{sf}}/\lambda \rangle$ and can be larger than $j_{e}/e$
(a special case, however, is that of a P configuration of a
symmetric structure, for which $m^{P}_{Cu}=0$). Enhancing the spin
accumulation and increasing its ratio to the current $j_{e}$ is
certainly the most promising way to reduce the critical current,
for example with materials in which a higher spin accumulation
splitting can be expected (magnetic semiconductors ?). This
dependence on SDL and "amplification" is also taken into account
in the model of Stiles and Zangwill~[11,12] or Kovalev \textit{et
al.}~[15], and in recent calculations of Slonczewski~[10]. This
"amplification" also turns out in the model of Shpiro \textit{et
al.}~[14] for the opposite limit of non-interfacial transfer. The
main difference between the two limits is the equal importance of
the terms $\mathbf{M_{1}\wedge (M_{1}\wedge M_{2})}$ and
$\mathbf{M_{1}\wedge M_{2}}$ in the torque of Shpiro \textit{et
al.}~[14]. We will see below that the experimental critical line
diagram of Fig.2 indicates a largely predominant
$\mathbf{M_{1}\wedge (M_{1}\wedge M_{2})}$ torque term.

\vspace{0.1in} \textbf{III. Discussion and Conclusion}
\vspace{0.1in}

Our expression of the torque, Eq.(\ref{torque2}), can be applied
to the interpretation of the experimental results.

(a) If the torque of Eq.(\ref{torque2}) is written as
$G^{P(AP)}j_{e}\times \mathbf{M_{1}\wedge(M_{1}\wedge M_{2})}$
and, when the excitation can only be an uniform precession, the
critical currents at zero field are expressed as ~[3,17,18]:
\begin{eqnarray}
j_{C}^{P\rightarrow AP}=-\frac{\alpha
\gamma_{0}}{G^{P}}(H_{an}+2\pi M) \nonumber \\
j_{C}^{AP\rightarrow P}=\frac{\alpha
\gamma_{0}}{G^{AP}}(H_{an}+2\pi M) \label{critcurr}
\end{eqnarray}
where $\alpha$ is the Gilbert coefficient, $H_{\mathrm{an}}$ is
the anisotropy field and $M$ the magnetization. By using
experimental data (interface resistances, interface spin memory
loss coefficient, SDL, etc) from CPP-GMR experiments~[21,22] to
calculate the spin accumulation in the Co/Cu/Co trilayer  and then
$\mathbf{\Gamma}^{P (AP)}$ and $G^{P(AP)}$ from
Eq.(\ref{torque2}), we obtain a reasonable agreement with our
experiments: $j_{C}^{P\rightarrow AP} = -~2.8\times 10^{7}$
A/cm$^{2}$ (exp.: $-~1.25\times 10^{7}$A/cm$^{2}$) and
$j_{C}^{AP\rightarrow P}= + ~1.05\times 10^{7}$ A/cm$^{2}$ (exp.:
$ + ~1.17\times 10^{7}$ A/cm$^{2}$)~\cite{remark}.

What can be also predicted for the critical currents is : i) their
proportionality to the thickness of the thin magnetic layer (this
follows from the assumption of interfacial transfer and has been
already observed~[3]); ii) their decrease as the thickness of the
thick magnetic layer increases, with saturation at a minimum level
when the thickness exceeds the SDL (60 nm in Co at low
temperatures, for example); iii) their increase (at the scale of
the mean free path in the spacer) when the spacer thickness
increases; iv) their definite dependence on the SDL in the layers
and leads.

(b) In finite applied field, a diagram of the type of Fig.~2, with
a crossover between the two regimes around $H = H_{\mathrm{an}}$,
is expected for a torque of the form $\mathbf{M_{1}\wedge
(M_{1}\wedge M_{2})}$. The equations of the critical lines and a
fit with experimental data has been presented elsewhere~[18].The
diagram expected for a torque $\mathbf{M_{1}\wedge M_{2}}$ does
not include a zone where both the P and AP configurations are
unstable (\textit{regime B} with progressive and reversible
transition) and cannot be fitted with the experiments on Co/Cu/Co
trilayers.

Although the spin transfer effect begins to be better understood,
the possibility of reducing sufficiently the critical currents for
practical applications is still a pending question. For
conventional ferromagnetic metals (Co, etc) and from numerical
applications of the model of this paper~[16], some reduction seems
possible but probably by less than an order of magnitude. As we
have pointed out, a stronger reduction might be obtained with
other types of magnetic materials permitting higher spin
accumulations. On the other hand, another type of spin transfer
effect is the current-induced domain wall motion~[2].~According to
recent experimental results of domain wall motion with relatively
small current densities~[24], this should be also a promising way
for current-induced switching.


\begin{thebibliography}{10}

\bibitem{slonczewski}
J. Slonczewski, J. Magn. Magn. Mat. {\bf 159},  L1  (1996)

\bibitem{berger}
L. Berger, J. Appl. Phys. {\bf 71}, 2721  (1992); Phys. Rev. B
{\bf 54}, 9353  (1996).

\bibitem{katine}
J.A. Katine, Phys. Rev. Lett. {\bf 84}, 3149  (2000); F.J. Albert
{\it et~al.} Appl. Phys. Lett. {\bf 77}, 3809  (2000).

\bibitem{sun}
J.Z. Sun {\it et~al.}, Appl. Phys. Lett. {\bf81}, 2202 (2002).

\bibitem{grollier}
J. Grollier {\it et~al.}, Appl. Phys. Lett. {\bf 78}, 3663 (2001).

\bibitem{urazhdin}
S. Urazdhin {\it et~al.} cond-mat/0303149.

\bibitem{wegrove}
J.E. Wegrove {\it et~al.}, Europhys. Lett. {\bf 45}, 626 (1999).

\bibitem{tsoi}
M. Tsoi {\it et~al.}, Phys. Rev. Lett. {\bf 80}, 4281  (1998).

\bibitem{waintal}
X. Waintal {\it et~al.}, Phys. Rev. B {\bf 62}, 12317 (2000).

\bibitem{bazaliy}
J. Slonczewski, J. Magn. Magn. Mat. {\bf 247}, 324 (2002); L.
Berger, J. Appl. Phys. {\bf 91}, 6795  (2002).

\bibitem{stiles1}
M. Stiles, A. Zangwill, Phys. Rev. B {\bf 66}, 01440 (2002).

\bibitem{stiles2}
M. Stiles, A. Zangwill, J. Appl. Phys. {\bf 91}, 6812 (2002).

\bibitem{zhang}
S. Zhang, P.M. Levy, A. Fert, Phys. Rev. Lett. {\bf 88}, 236601
(2002)

\bibitem{shpiro}
A. Shpiro, P.M. Levy, S. Zhang, Phys. Rev. B {\bf 67}, 104430
(2003).

\bibitem{xia}
K. Xia {\it et~al.}, Phys. Rev. B {\bf 65}, 220401 (2002); A.
Kovalev {\it et~al.}, \textit{ibid}. {\bf 66}, 224424 (2002).

\bibitem{fert}
A. Fert {\it et~al.}, to be published

\bibitem{sun2}
J.Z. Sun, Phys. Rev. B {\bf 62}, 570 (2000).

\bibitem{grollier2}
J. Grollier {\it et~al.}, Phys. Rev. B, in press.

\bibitem{miltat}
J. Miltat {\it et~al.}, J. Appl. Phys. {\bf 89}, 6982 (2001).

\bibitem{valetfert}
T. Valet and A. Fert, Phys. Rev. B {\bf 48}, 7099 (1993).

\bibitem{bass}
J. Bass and W.P. Pratt, J. Magn. Magn. Mat. {\bf 200}, 274 (1999)
; A. Fert and L. Piraux, ibidem, 338.

\bibitem{park}
W. Park {\it et~al.}, Phys. Rev. B {\bf 62}, 1178 (2000).

\bibitem{remark}
These values are slightly different from those of Ref[18] where
the interface spin memory loss is not taken into account in the
calculation.

\bibitem{gan}
J. Grollier {\it et~al.}, J. Appl. Phys. {\bf 92}, 4825 (2002) and
to be published.

\end{thebibliography}
\end{document}